# Softening of Vibrational Modes and Anharmonicity Induced Thermal Conductivity Reduction in a-Si:H at High Temperatures

Zhuo Chen[1,2], Yuejin Yuan[1*], Yanzhou Wang[3], Penghua Ying[4], Shouhang Li[5], Cheng Shao[6], Wenyang Ding[7], Gang Zhang[2*], Meng An[7*]

[1] College of Mechanical and Electrical Engineering, Shaanxi University of Science and Technology, Xi'an, Shaanxi 710049, China

[2] Yangtze Delta Region, Academy of Beijing Institute of Technology, JiaXing, China

[3] Department of Applied Physics, QTF Center of Excellence, Aalto University, FIN-00076 Aalto, Espoo, Finland

[4] Department of Physical Chemistry, School of Chemistry, Tel Aviv University, Tel Aviv, 6997801, Israel

[5] Centre de Nanosciences et de Nanotechnologies, CNRS, Université Paris-Saclay, 10 Boulevard Thomas Gobert, 91120 Palaiseau, France

[6] Thermal Science Research Center, Shandong Institute of Advanced Technology, Jinan, Shandong 250103, China

[7] Department of Mechanical Engineering, The University of Tokyo, 7-3-1 Hongo, Bunkyo, Tokyo 113-8656, Japan.

Corresponding author: yuanyj@sust.edu.cn (Y. Y. J.);

gangzhang2006@gmail.com (G. Z.)

anmeng@photon.t.u-tokyo.ac.jp (M. A.)

**Abstract**

Hydrogenated amorphous silicon (a-Si:H) has garnered considerable attention in the semiconductor industry, particularly for its use in solar cells and passivation layers for high-performance silicon solar cells, owing to its exceptional photoelectric properties and scalable manufacturing processes. A comprehensive understanding of thermal transport mechanism in a-Si:H is essential for optimizing thermal management and ensuring the reliable operation of these devices. In this study, we developed a neuroevolution machine learning potential based on first-principles calculations of energy, forces, and virial, which enables accurate modeling of interatomic interactions in both a-Si:H and a-Si systems. Using the homogeneous nonequilibrium molecular dynamics (HNEMD) method, we systematically investigated the thermal conductivity of a-Si:H and a-Si across a temperature range of 300-1000 K and hydrogen concentrations ranging from 6 to 12 at%. Our simulation results found that thermal conductivity of a-Si:H with 12 at% hydrogen was significantly reduced by 12% compared to that of a-Si at 300 K. We analyzed the spectral thermal conductivity, vibrational density of states and lifetimes of vibrational modes, and revealed the softening of vibrational modes and anharmonicity effects contribute to the reduction of thermal conductivity as temperature and hydrogen concentration increase. Furthermore, the influence of hydrogen concentration and temperature on diffuson and propagon contribution to thermal conductivity of a-Si:H was revealed. This study provides valuable insights for developing thermal management strategies in silicon-based semiconducting devices and advances the understanding of thermal transport in amorphous systems.

# 1. Introduction

The thermal properties of amorphous solids are of fundamental interest due to the strong localization of vibrational modes and the suppression of the length scale of scattering. Heat conduction in crystalline solids with complete periodicity is well understood in terms of mode-dependent phonon transport properties. Peierls[1] developed the Boltzmann transport equation (BTE) for phonon transport can predict the heat conduction properties of various nanostructured materials. The wave and particle duality of vibration modes can be well described with the framework of lattice dynamics[2] and Boltzmann transport equation[3]. However, the established Boltzmann phonon transport theory assumes a well-defined group velocity. In amorphous solids, only a small fraction of vibration modes with low frequencies possesses a crystalline-like group velocity. Allen and Feldman[4] classified the vibrational modes in amorphous materials into three categories: propagons, diffusons, and locons. Propagons are propagating and delocalized phonon-like plane waves that typically possess long wavelengths compared to the interatomic spacing. Diffusons are modes that scatter over a distance less than the interatomic distance and thus transport heat as a random-walk. Locons are non-propagating and localized modes that are unable to transport heat in harmonic solids. These categorizations have been used to understand the thermal transport in various amorphous solid materials[5, 6].

Recently, hydrogenated amorphous silicon (a-Si:H) have attracted significant attention across various fields, including solar cells, thin film transistor and infrared detectors. This is due to its enhanced electrical properties, which arise from the termination of the dangling bonds and modification of short-range order[7] compared with conventional amorphous silicon. For example, past study reported that a-Si:H-based solar cells achieved the ultrahigh conversion efficiency of photovoltaic devices[8]. In these applications, devices generally operate at high temperatures, where the thermal transport properties of the materials, including a-Si:H, play a critical role in determining their reliability and lifespan[9, 10]. To achieve optimal device performance, it is essential to understand the thermal transport properties of a-Si:H with varying hydrogen concentrations, particularly at high temperatures. The extended 3ω measurement technique [11] was used to measure the thermal conductivity of sputtered a-Si:H thin film for a hydrogen concentrations ranging within 1-20%. Moreover, the time-domain thermoreflectance (TDTR) method was developed to measure [12] the thermal conductivity of a-Si:H films synthesized from plasma-enhanced chemical vapor deposition method, where the results was found that the contribution of diffusons and propagons to thermal conductivity is largely suppressed when the introduction concentration of hydrogen atoms is 25.4 wt%. In addition, the thermal conductivity of amorphous materials was modulated via the structural features from transmission electron microscopy (SEM) of experimental samples[13], nanoporous structures[6], microstructure[14, 15], and superlattice[16]. However, accurately predicting or measuring the thermal transport properties of *a-Si*:H remains challenging due to the complexities of its interatomic interaction in large-scale molecular dynamics (MD) simulations and its structural disorder.

In recent years, machine learning potentials have revolutionized MD simulations for studying thermal transport properties in complex and amorphous materials. These advancements offer significant benefits, including enhanced computation efficiency and the ability to capture all orders of anharmonicity. For example, Gaussian approximation potential (GAP) [17] was developed for predict the interatomic force of crystalline and a-Si, where the calculated thermal conductivity of crystalline silicon and a-Si are 121 W/m-K and 1.4 W/m-K at room temperature, respectively, aligning well with experimental measurements and first-principle calculations. To balance the tradeoff between the computational efficiency and accuracy, neuroevolution potential (NEP) [18] was developed to describe the interatomic interactions in a-Si with a large-scale simulation system of 64000 atoms. Remarkably, the NEP not only achieved high accuracy in thermal conductivity calculation but also demonstrated computational cost comparable to, or even than those of empirical potentials. Subsequent studies have developed additional machine learning potentials for amorphous solids, further highlighting the advantages of NEP in term of the accuracy and computational efficiency[19] across various amorphous materials, including a-Si[18], a-HfO$_2$[20], a-SiO$_2$[21]. Despite these advances, accurately predicting the thermal transport properties of a-Si:H[12] and unraveling the thermal transport mechanism at high temperatures remains unresolved challenges due to the increased structure instability and disorder by the incorporation of light-mass hydrogen atoms.

In this work, we developed the neuroevolution potential of a-Si:H with the temperatures from 100 to 1200 K and hydrogen concentrations from 6 to 12 at% and investigated the influence of hydrogen concentration on the thermal conductivity of a-Si at various temperatures. The interatomic interactions in a-Si:H and a-Si in previously developed a GAP potential for a-Si:H[22] was used to construct NEP, which was fitted based on the separable natural evolution strategy (SNES)[23]. The radial distribution function (RDF) and static structural factors (SF) of *a-Si*:H and *a-Si* were calculated to validate the reliability of NEP. Homogeneous nonequilibrium MD (HNEMD) method [24] was used to calculate the thermal conductivity of *a-Si*:H and *a-Si* over a temperature range of 300 to 1000 K, as well as H concentrations ranging from 6 to 12 at%. Additionally, the thermal transport machanism of a-Si and a-Si:H at high temperatures was systematically analyzed by the vibrational density of states (VDOS) and vibrational modes using normal-mode decomposition (NMD) method.

## 2. Training and validating of NEPs for *a-Si* and *a-Si*:H
### 2.1 The training of NEPs

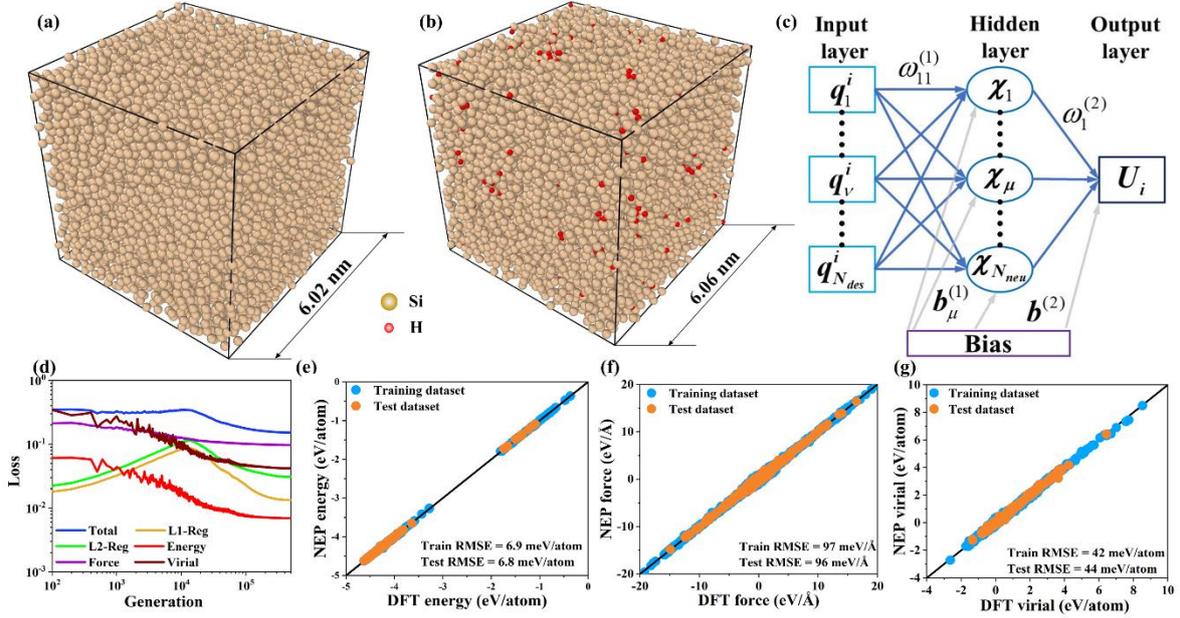

Figure 1. The MD snapshots of (a) a-Si and (b) a-Si:H with 6 at% after annealing at 300 K, where the yellow (red) spheres are Si and H atoms, respectively. (c) Example of a standard neural network with a single hidden layer. (d) Evolution of the loss function of L1 and L2 regularization, the energy RMSE (eV/atom), the force RMSE (eV/Å), and the virial RMSE (eV/atom). (e)-(g) The comparison between the NEP predictions and DFT reference data of energy, force, and virial for the training and testing data sets. The lines in (e)-(g) are a guide for the eyes.

The NEP model is a machine learning potential (MLP) based on the feedforward neural network, is employed to accurately describe the interatomic interactions. One hidden layer is used in NEP, as shown in the Fig. 1(c). The neuron in the output layer is the energy $U_i$ of atom $i$, which depends on the values of the $N_{des}$ input neurons also called descriptor vector. In the middle is a hidden layer with $N_{neu}$ neurons. The blue arrows represent the weight parameters and gray are the bias parameters, where the bias adjusts the activation functions' region of nonlinearity. To compare the thermal transport features of a-Si and a-Si:H, two NEP potentials are trained independently. For a-Si, we retrain a NEP based the comprehensive training database of GAP in previous study[25]. This training set contains various silicon structures, including the liquid , crystalline , and amorphous atates. All of NEP models are trained by the GPUMD package[26] (version 3.8, NEP4). The computational speed of $3\times10^6$ atom-step per second for a-Si contains 10648 atoms by using one Tesla V100 GPU, which is about 1000 times faster than GAP using 72 Xeon-Gold 6240 CPU cores[17]. For *a-Si*, previous work[18] has trained a NEP model with this training set and has been validated to be reliable. In this work, we use the same training set and parameters to get a NEP again. In Fig. S1(a) we show the trend of convergence of each item in the loss function during the training process. We can found it the curves of energy, force, and virial tends to be horizontal within the number of generations. In Fig. S1(b)–(d) we compare the energy, force, and virial predicted by the NEP and DFT calculations for the training and testing datasets. The train RSMEs of energy, force, and virial are 6.9 meV/atom, 97.0 meV/Å, and 42.0 meV/atom, respectively. As a comparison, the corrresponding RMSEs for the hold-out testing date from GAP[25] are 7.8 meV/atom and 93 meV/Å (virial RMSE is not available), respectively. The accuracy achieved exhibit comparable accuracy to that reported by the GAP model. But the NEP has the efficiency far exceeds the GAP.

For a-Si:H, the training dataset consists of 390 cells collected from another GAP[22]. The doped structures of the training set were gathered by adding H to supercells of pure Si and other structures including a-Si, liquid Si, diamond Si, and amorphous/crystalline Si interface structures. For the liquid and amorphous phases, the atomic concentration of incorporated H was within the range of 6 to 12 at%, which are considered in our work. The parameters used for training the NEP are same as a-Si. The NEP is rigorously evaluated through comparisons of the energy values and atomic forces between DFT calculations and NEP predictions, demonstrating excellent quantitative agreement, as shown in Fig. 1(d)-(g). The energy, force and virial RMSEs for the training dataset are 3.6 meV/atom, 123 meV/Å, and 21 meV/atom. As a comparison, the energy and force RMSEs for the training set from GAP for a-Si:H are 4 meV/atom and 200 meV/Å. Here, the accuracy of the NEP even surpasses that of the GAP. The NEP model's notable speed and precision lay the groundwork for an extensive investigation into heat transport in a-Si:H. Subsequently, we will demonstrate that NEP also exhibits a high level of precision in forecasting structural properties.

## 2.2 Atomic structure and validation

All amorphous structures are obtained by the classical MD simulations with a melt-quench-anneal process (see Fig. S2). All the MD simulations are performed using the GPUMD package with the NEPs[26] (version 3.8). We get a cubic crystal of diamond silicon from Material Project website [27] as the initial structure. Here we set the conventional cell of Si as a unit cell and the lattice constant as 5.45 Å. Taking into account computational expenses and size effects, we ultimately selected a $11 \times 11 \times 11$ supercell consisting of 10648 Si atoms as the structure for MD simulations and a-Si:H is obtained by randomly adding H atoms to a-Si using the Atomic Simulation Environment (ASE) package[28]. Subsequently, the melt-quench-anneal strategy is implemented, the diagram of simulation process is shown in the Fig. S2. We use the NVT (the canonical ensemble) with the Bussi-Donadio-Parrinello[29] method during the melt-quench process and NPT the (isothermal-isobaric ensemble) at zero pressure with stochastic cell rescaling method[30] is used in the annealing. The snapshots of a-Si and a-Si:H at 300 K after the melt-quench-anneal process are presented in Fig. 1 (a)(b). The structure of a-Si:H is obtained through MD-NEP with a mass density of approximately 2.21 g/cm$^3$, consistent with PECVD films[31]. More structural properties are showed in Fig. 2, the radial distribution function (RDF) g(r) and structral factor (SF) S(q) are calculated to validate the NEP's reliability.

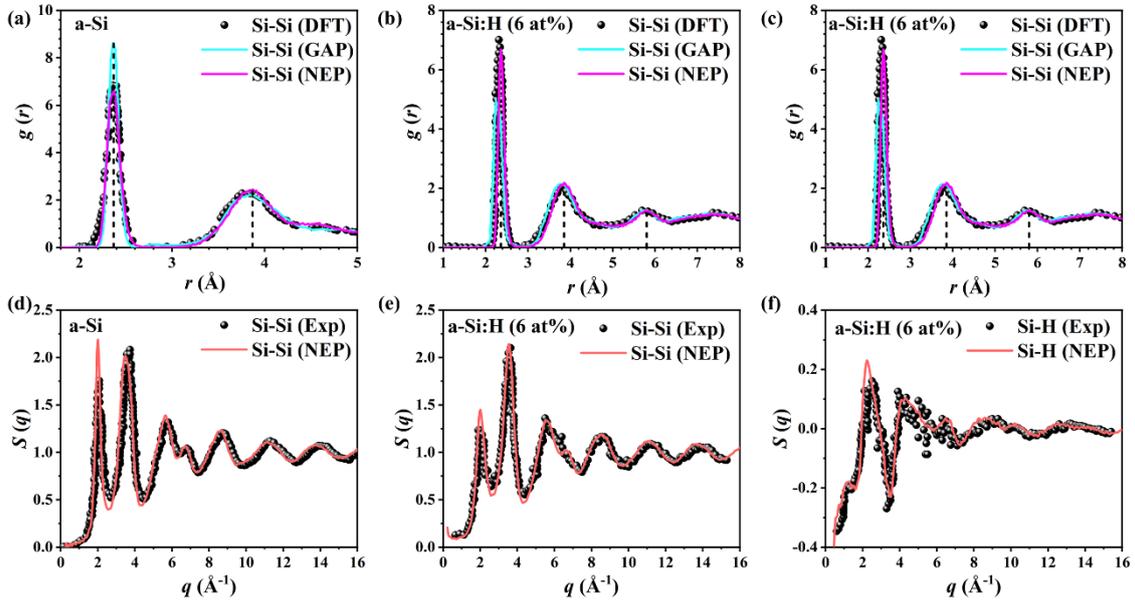

Figure 2. (a) Radial distribution function g(r) of a-Si at 300 K. GAP[25] and experimental results[32] are shown for comparison. Radial distribution function g(r) of (b) Si-Si and (c) Si-H in a-Si:H with 6 at% at 300 K. GAP force field[22] and DFT calculations results[33] are shown for comparison. The peaks of g(r) are marked with the vertical dash lines. The structure factors S(q) of (d) Si-Si in a-Si, (e) Si-Si and (f) Si-H in a-Si:H, where experimental results[32, 34] are shown for comparison.

A distinctive feature of the amorphous structure lies in their fundamental deviation from crystalline arrangements, most notably manifested by the complete lack of the long-range order in crystalline structure, while it still exhibits short-range and medium-range ordering[35]. After generating the a-Si and a-Si:H samples, the RDF were computed at 300 K to characterize the short-range ordered bond motifs in the system. The result of a-Si is shown in Fig. 2(a). The key features of g(r) of a-Si located at about 2.37 Å and 3.87 Å, both results demonstrated remarkable consistency with the reference data g(r) of the experimental result[32]. While the GAP model[25] has a higher distribution in the first peak. Unlike pure a-Si, a-Si:H (6 at%) encompasses the interaction between Si and H. Therefore, results are given for the Si-Si and the Si-H partial RDF g(r). The RDF analysis reveals distinct structural features for both Si-Si and Si-H atomic pairs. As shown in Fig. 2(b)(c), for the Si-Si pair g(r), three prominent peaks are observed at approximately 2.36 Å, 3.85 Å, and 5.8 Å, which correspond to the first, second, and third coordination shells, respectively. Similarly, the Si-H pair g(r) exhibits four characteristic peaks located at about 1.5 Å, 3.16 Å, 5.0 Å, and 6.6 Å. Remarkably, the Si-Si and the Si-H partial RDF of a-Si:H, produced with our NEP, both achieved more significant agreement with the reference g(r) of the DFT calculations[33] than GAP[22], both peak locations and heights. Additionally, the medium-range ordering, another typical characteristic of amorphous structures, is often captured experimentally by the static SF S(q) of X-ray diffraction[36]. In experiments, SF is computed computed by diffraction date, followed by deriving RDF via the Fourier transform. Conversely, MD simulations first ascertain RDF, subsequently obtaining SF through a transformation process:

$$S(q) = 1 + 4\pi\rho_0 \int_0^\infty r^2 [g(r) - 1] \frac{sin(qr)}{qr} dr. \qquad (4)$$

where $q$ is wave vector, $q = \frac{4\pi \sin\theta}{\lambda}$, $\lambda$ is the wavelength of the incident wave. $r$ represents the radius of the atom's RDF. $\rho_0$ is the average atomic number density in the system.

For the comparison with diffraction experiments[32, 34] (there is no reference data for GAP), we also calculate the static SFs using the ISAACS package[37]. As shown in Fig. 2(d), the computed S(q) of a-Si qualitatively reproduces the experimental features[32], for the height and position of the diffraction peaks. In the same, we can find a very good agreement of SFs with that of the experimental x-ray diffraction experiments[34] for a-Si:H in Fig. 2(e)(f). In summary, the accuracy of the structure obtained by MD-NEP has been verified and is sufficient for subsequent thermal transport calculations.

## 2.4 Thermal conductivity calculation method

Homogeneous non-equilibrium molecular dynamics (HNEMD)[24] is adopted to calcualte the thermal condcutivity of a-Si and a-Si:H, which was sucessfully reproduced the thermal conductivity of various amorphous and crytalline materials[18, 20, 21]. In our simulations, $F_e$ = 2 × 10$^{-4}$ /Å in previous report[18] is adopted to calculate the thermal conductivity of a-Si samples. For a-Si:H, a value of $F_e$ is not enough for all samples due to its high sensitivity to variations in temperature and hydrogen atom concentrations. Therefore, we carried out extensive tests to determine the parameter for each case. By testing, 0.001 /Å to 0.0025 /Å are choosed to the target temperature (300 to 1000 K) and the samples with 6 ~ 12 at%. The thermal conductivity is calculated at zero pressure with the HNEMD method after the melt-quench-anneal process. We first equilibrate each system for 0.1 ns and then apply the external force for 8 ns, a time step of 0.5 fs is used. For all the systems, we perform three independent calculations and average the results with a proper estimation of the statistical error (as shown in Fig. S4 and S5).

**2.5 Vibrational mode lifetime calculation**

The lifetimes of vibrational modes are computed based on normal mode decomposition (NMD) method[38-40]. To perform the NMD, the equilibrium configurations of a-Si and a-Si:H are firstly obtained through melt-quench-anneal. After annealing, the system is equilibrated with a time step of 0.5 fs in a NVT ensemble for 10 ps at the targeted temperature, and then the velocities of each atom are dumped every 10 fs in NVT ensemble during a time span of 0.5 ns for postprocessing, all of MD trajectories are dumped by LAMMPS[41] that built NEP interface. Force constants are calculated by the General Utility Lattice Program[42] based on the equilibrium configurations of *a-Si* and *a-Si*:H. Then, the anharmonic vibrational frequency and lifetime are obtained through NMD method using the DynaPhoPy code[43]. The NMD-predicted lifetimes are plotted in Fig. 4(c) and the value of a-Si is broadly consistent with previous research[39].

**3. Results and discussions**

**3.1 Thermal conductivity of a-Si and a-Si:H**

Firstly, the thermal conductivities of a-Si:H with various hydrogen concentrations were calculated utilizing HNEMD method, as shown in Fig. 3(a). It is clearly obseved that the thermal condcutivity of a-Si:H decreases from 1.7 to 1.0 W/m-K as the hydrogen concentration increases from 0 to 12 at%. The trend aligns closely with previous experiments results[11, 12], further validating our calculation results. In other words, our trained NEP has been sucessfully predicted the thermal condcutivity of a-Si:H, demonstrating its reliability and accuracy in capturing the influence of hydrogen concentration on thermal transport properties. The slightly difference between our caluclation with experimental measusrments may stem from the various quality of a-Si:H samples from plasma-asisted chemical vapor deposition (CVD) method[11] and and the plasma-enhanced CVD[12]. In the classical molecular dynamics simulations, vibrational modes follow the Maxwell-Boltzmann statistics, indicating each vibrational mode contriubtes to the heat capacity. However, in the quatunum statistical mechanics, only the vibrational modes with the frequency cutoff below $k_B T \leq \hbar \omega_0$, with $\omega_0$ the vibrational frequency contribute to heat capacity according to Bose-Einstein statistics. One quantum-correction method was used to calculate the quantum modal heat capacity based on the spectral thermal conductivity and classical modal heat capacity[18]:

$$\kappa^q(\omega, T) = \kappa(\omega, T) \frac{x^2 e^x}{(e^x-1)^2}, \quad (15)$$

$$\kappa^q(T) = \int_0^\infty \frac{d\omega}{2\pi} \kappa^q(\omega, T). \quad (16)$$

where $x = \hbar\omega/k_B T$, $\hbar$ is the reduced Plank constant and $k_B$ is the Boltzmann constant. $\kappa(\omega, T)$ is the classical spectral thermal conductivity and the total quantum corrected thermal conductivity $\kappa^q(T)$ is obtained as an integral of $\kappa^q(\omega, T)$ over the vibrational frequency. In Fig. 3(a), all the quantum corrected thermal conductivities of a-Si:H with various hydrogen concentrations at room temperature are slightly smaller than that of classcial MD simulations. Because vibrational modes at room temperature in classical molecular dynamics simulations are completed activated, which overestimates the number of vibrational modes at high-frequency range in quantum statistical mechanics.

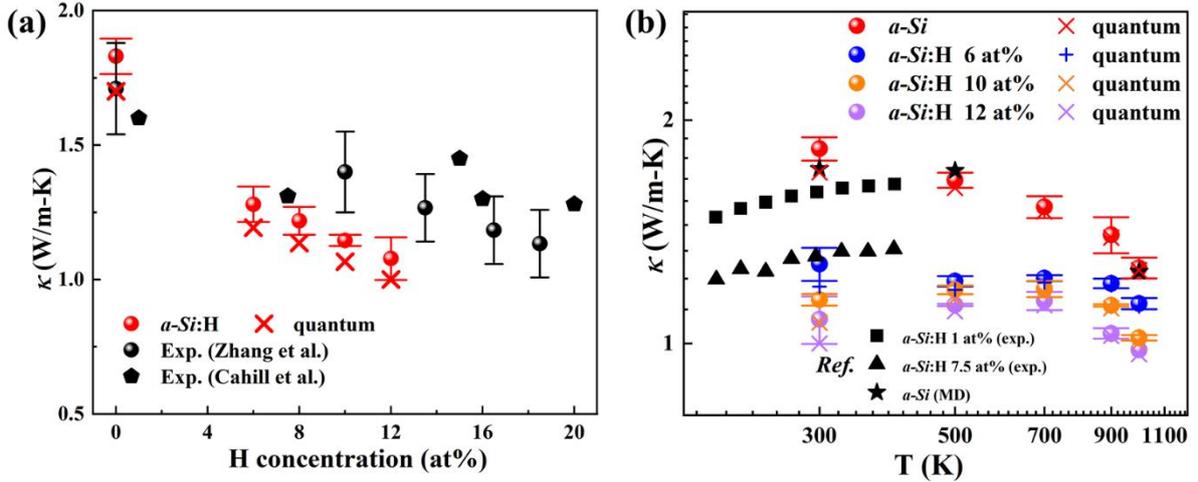

Figure 3. (a) Hydrogen concentration-dependent thermal conductivity for *a-Si*:H as compared to the experiment values by Cahill et al.[11] and Zhang et al.[12]. (b) The temperature-dependent thermal conductivity for a-Si and a-Si:H with 6 at%, 10 at% and 12 at% hydrogen concentration from our calculations as compared to the experiment values by Cahill et al.[11]. The another HNEMD result by Wang et al.[18] based on NEP are also shown for comparison.

Moreoever, the thermal conductivity stabilty at various temperatures plays a central role in the performance and lifetime of devices. The temperature dependence of thermal conductivity in a-Si:H was also investigated and the thermal conductivity of a-Si and a-Si:H with 6, 10 and 12 at% concentration are shown in Fig. 3(b) when the temperature ranges from 300 to 1000 K. To validate the calculated thermal conductivity realiability of a-Si under different temperatures, the measured thermal conductivity of a-Si:H with 1 and 7.5 at% under around 300 K are plotted in Fig. 3(b). It is clearly observed that our MD results shows a high degree of accuracy in both the value of thermal conductivity or their temperature dependnce with that in experiment. Moreover, both the thermal conductivity of a-Si and a-Si:H decrease with temperature ranging from 300 to 1000 K. Meanwhile, the significant quantum effect of thermal conductivity in a-Si have been observed at low temperature, where the quantum-corrected thermal condcutivity of a-Si:H is decreased by 7% at 300 K compared with that from classical MD simulations. However, the quantum effect of thermal conductivity is negligible due to vibrational modes at medium and high temperatures in quantum statistics, approachinng to the completely activation of vibrational modes in classical Maxwell-Boltzmann statistics.

**3.2 Mechanisms of hydrogen concentration-dependent thermal conductivity of *a-Si*:H**

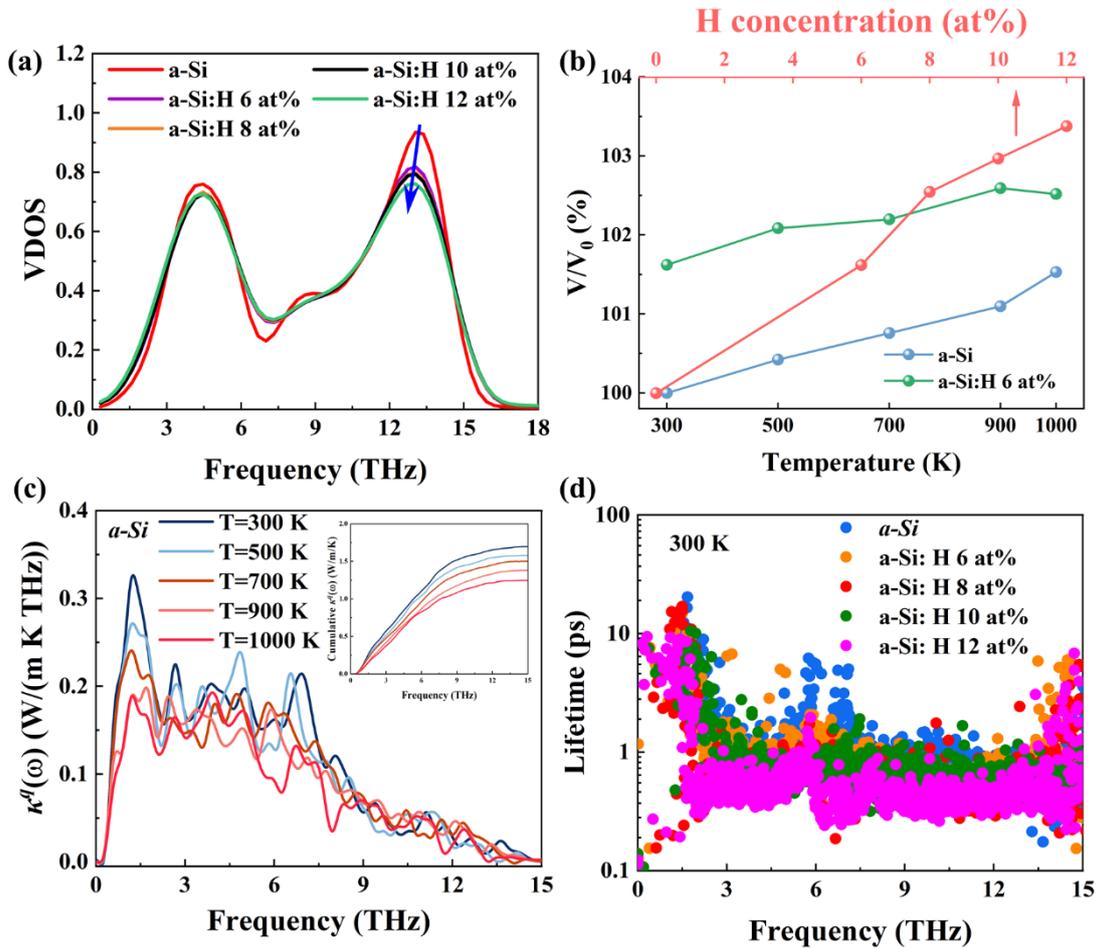

Figure 4. (a) The vibration density of states for silicon atoms in a-Si:H with the hydrogen concentrations from 6 at% to 12 at%. (b) The ratio of volume for a-Si:H at the temperatures from 300 K to 1000 K and volume for a-Si at 300 K. (c) The quantum-corrected spectral thermal conductivity and (d) lifetimes of vibrational modes for a-Si:H at 300 K with hydrogen concentrations from 0 at% to 12 at%.

Previous studies have demonstrated that the mechanism of thermal conductivity regulation for amorphous materials by heteroatom doping is generally attributed to three aspects in the following: changes in heat capacity due to mass differences, variation in sound velocity induced by voids and enhanced scattering of heat carriers. For a-Si:H, it can be deduced that the hydrogenation of amorphous silicon results in a reduced heat capacity due to the decrease of mass density. Moreover, the vibrational density of states (VDOS) for silicon atoms, reflecting the vibration modes properties of solid materials, were calculated in Fig. 4(a) to analyze the regulation mechanism of vibration modes in a-Si:H by hydrogenation. As shown in Fig. 4(a), the vibration density of states of a-Si:H is predominately characterized by two distinct regions: low-frequency vibrational modes centered around 4.5 THz and high-frequency vibrational modes near 13.5 THz. Compared with a-Si, the amplitude of both peaks in VDOS for a-Si:H are significantly suppressed, indicating that the hydrogenation effectively reduces the thermal conductivity by restricting the vibrational modes within a-Si. Interestingly, high-frequency vibrational modes near 13.5 THz exhibits a redshift, indicating a softening of vibrational modes caused from void formation resulting from the reduced coordination number of silicon atoms in a-Si:H. To quantify such an effect, the volumes of simulation cell in MD simulation at 300 K for different hydrogen concentrations are extracted, as shown in Fig. 4(b). The volume expansion ratio ($V/V_0$) of a-Si:H at varying hydrogen concentrations was employed to quantify the volumetric changes in the MD simulation cell induced by hydrogenation. Interestingly, the volume expansion ratio exhibits a notice increase with rising hydrogen concentration. In previous work[20], Zhang et al. demonstrated that the softening of vibrational modes caused by volume expansion leads to the reduced thermal conductivity, as evaluated using the quasiharmonic Green-Kubo (QHGK) method. Building on the aforementioned analysis, it can be inferred that the hydrogenation of a-Si induces vibrational modes softening, altering the system volume of MD simulations, further contributes to the reduction of thermal conductivity.

Furthermore, we calculated the quantum-corrected spectral thermal conductivity of a-Si:H under various hydrogen concentrations at 300 K, as shown in Fig. 4(c). Clearly, the thermal conductivity of a-Si:H across all hydrogen concentrations are dominately contributed by vibrational modes below 15 THz. Specifically, the reduction of thermal conductivity by hydrogen doping predominantly occurs in the frequency from 3 to 9 THz. In addition to the harmonic properties, the anharmonicity of vibrational modes in a-Si:H are also evaluated by the lifetime, as shown in Fig. 4(d). Notably, the lifetime of vibrational modes significantly decreases with increasing hydrogen concentration. However, the lifetime of vibrational modes below 3 THz remains largely unaffected, indicating that low-frequency vibrational modes are less sensitive to changes in hydrogen concentration.

## 3.3 Mechanisms of temperature-dependent thermal conductivity of a-Si:H

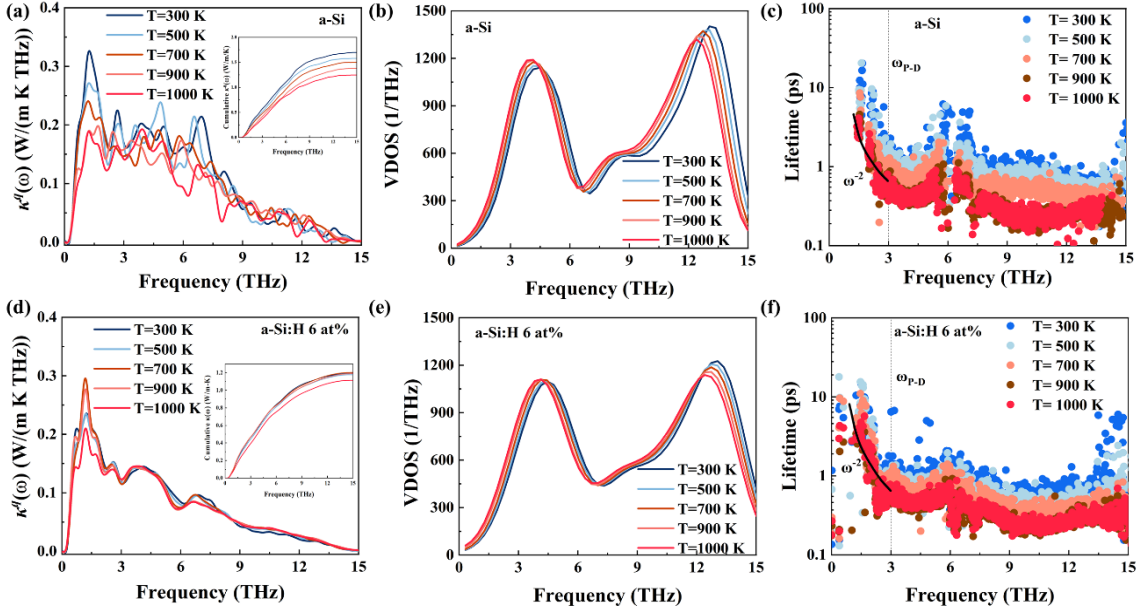

Figure 5. The quantum-corrected spectral thermal conductivity of (a) a-Si and (d) a-Si:H with 6 at% at these temperatures from 300 K to 1000 K. The vibrational density of states of Si atoms in (b) a-Si and (e) a-Si:H with 6 at% at these temperatures from 300 K to 1000 K. The lifetimes of vibrational mode of (c) a-Si and (f) a-Si:H with 6 at% at these temperatures from 300 K to 1000 K.

In addition to the influence of hydrogenation on thermal conductivity of a-Si, we also explore the temperature dependence of thermal conductivity in a-Si and a-Si:H with 6 at%. We first analyzed the quantum-corrected spectral thermal conductivity of a-Si under temperature ranging from 300 to 1000 K in Fig. 5(a). It is found that the low-frequency vibrational modes of a-Si are significantly suppressed as temperature increases. Moreover, the calculated VDOS in Fig. 5(b) exhibits an obvious redshift at high temperatures, indicating the softening effect of vibrational modes and volume expansion in Fig. 4(b). The softened vibrational modes decrease their group velocity of vibrational modes, contributing to the reduced thermal conductivity. Furthermore, the temperature-induced anharmonicity was quantified by the lifetime of vibrational modes, as shown in Fig. 5(c). The Ioffe-Regel crossover frequency, $\omega_{P\text{-}D}$ was determined based on propagon with well-defined excitation due to the short-range and medium-range ordering and diffuson with no lattice dispersion in amorphous solids. In addition, the contribution of both propagons below 3 THz and diffusons above 3 THz decrease when the temperature increases from 300 to 1000 K.

In contrast, the quantum-corrected spectral thermal conductivity of a-Si:H with 6 at% under various temperature in Fig. 5(d) exhibits a distinct behavior. Specifically, the reduction of thermal conductivity in a-Si:H with increasing temperature is primarily attributed to the low-frequency vibrational modes below 3 THz. Moreover, the weaker softening phenomenon of vibrational modes in the VDOS is observed for a-Si:H compared with a-Si, which aligns with its less pronounced temperature dependence of thermal conductivity. In other words, the volume changes due to thermal expansion in a-Si:H at high temperature is less significant than that in a-Si. It is noteworthy that the softening of the vibrational mode, induced by the anharmonic effect at high temperatures, results in decreased thermal conductivity of a-Si and a-Si:H. Furthermore, the anharmonicity-induced vabrational modes in a-Si:H at high temperatures were analyzed by calculating the lifetimes of vabratonal modes, as shown in Fig. 5(f). Notably, the lifetime of propagon below 3 THz remains largely unchanged, whereas the lifetime of diffusons exhibits a significant decrease with increasing temperatures. The difference in lifetime between a-Si and a-Si:H can be attributed to hydrogenation, which induces void formation as a result of the reduced coordination of silicon atoms. The structural alternation significantly impacts the propagons associated with low-frequency vibrational modes, highlighting the role of hydrogenation in modifying the vibration dynamics and thermal transport properties of a-Si:H.

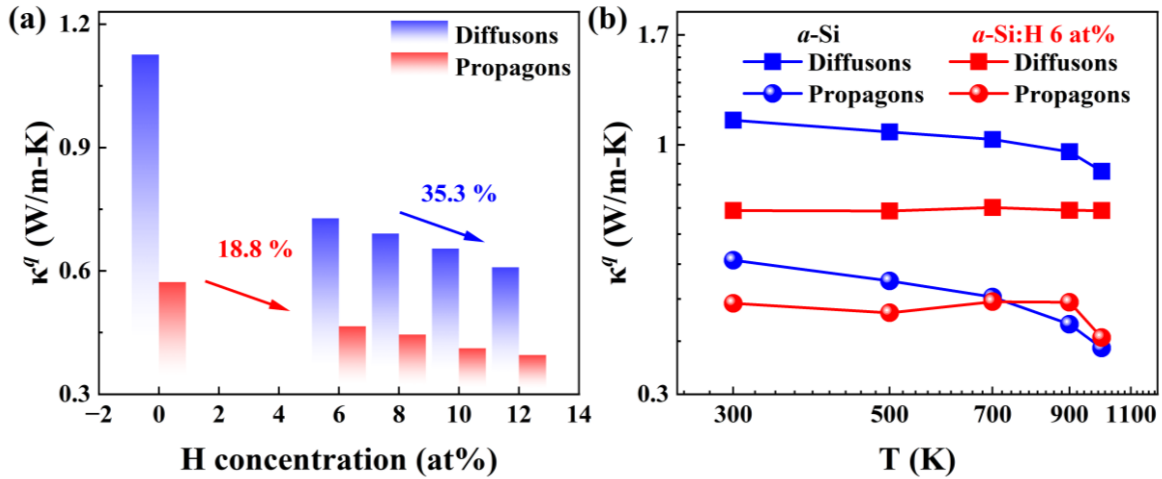

Figure 6 The diffusons and propagons contribution of thermal conductivity in (a) a-Si:H with different hydrogen concentrations at 300 K and (b) The temperature dependence of thermal conductivity of a-Si and a-Si:H 6 at% from diffusons and propagons contributions.

Furthermore, the contributions of diffusons and propagons to the thermal conductivity in a-Si:H with varying hydrogen concentrations were calculated, as illustrated in Fig. 6(a). The results clearly indicate that both diffusons and propagons exhibits reduced contributions to the thermal conductivity as the hydrogen concentration increases. Specifically, the contribution from propagons decreases by 18.8%, while that of diffusons decreases by 35.3% when the hydrogen concentration increases from 0 to 12 at%. This trend underscores the significant impact of hydrogen doping in altering the diffusons-mediated contribution to thermal conductivity in a-Si:H. Additionally, the temperature dependences of diffusons and propagons in a-Si and a-Si:H with 6 at% is presented in Fig. 6(b). In a-Si, both diffusons and propagons decreases with increasing temperature, exhibiting similar magnitudes of variations. However, in a-Si:H, the behavior differs: the contribution from diffusons remains nearly unchanged, whereas propagons exhibits a sharp decline with increasing temperature. These results highlight that the contribution of propagons is more sensitive to temperature compared to diffusons in both a-Si and a-Si:H.

## 4. Conclusions

In conclusion, this study investigated the impact of hydrogenation on the thermal conductivity of a-Si:H across temperatures ranging from 300 to 1000 K using the machine learning potential-based HNEMD method. The simulation results demonstrate that the thermal conductivity of a-Si:H decreases by 41% at 300 K as the hydrogen concentration increases from 0 to 12 at%. The underlying mechanism is attributed to hydrogenation-induced softening of vibrational modes, which alters the system volume in MD simulations and contributes significantly to the reduction in thermal conductivity. Additionally, the thermal conductivity of both a-Si and a-Si:H (6 at%) decreases with increasing temperature. Notably, the temperature dependence of thermal conductivity is more pronounced in a-Si compared to a-Si:H, primarily due to the less significant softening of vibrational modes in the hydrogenated material. Moreover, void formation in a-Si:H caused by hydrogenation enhances the scattering of vibrational modes below 3 THz. Furthermore, the interplay between vibrational mode softening and anharmonicity results in a decrease in the contributions of both diffusons and propagons in a-Si:H as the hydrogen concentration increases. Among these, the contribution of propagons is more significantly reduced in both a-Si and a-Si:H compared to diffusons. These findings provide valuable insights into the mechanisms governing thermal transport in a-Si:H and other amorphous materials, highlighting the role of hydrogenation in modulating their vibrational dynamics and thermal properties.

**Acknowledgements**

This work was supported by the National Natural Science Foundation of China (No. 52376063), the Hebei Key Laboratory of Low Carbon and High Efficiency Power Generation Technology Prevention Fund (Grant No. 2022-K03) and the China Postdoctoral Science Foundation (No. 2023MD744223).